\documentclass[aps,prl,twocolumn,superscriptaddress,showpacs]{revtex4-2}
\usepackage[T1]{fontenc}
\usepackage{graphicx}
\usepackage{xcolor}
\usepackage{dcolumn}
\usepackage{comment}
\usepackage{amsmath,amssymb,bbold,bm}
\usepackage{amsfonts}
\usepackage{float,latexsym}
\usepackage{multirow}
\usepackage{color}
\usepackage[normalem]{ulem}
\usepackage{physics}
\usepackage{tabularx}
\usepackage{array}
\usepackage[caption=false]{subfig}
\usepackage{siunitx}
\usepackage{gensymb}
\usepackage{pifont}
\usepackage{simplewick}
\usepackage{placeins}
\usepackage{mathtools}
\usepackage{longtable}
\usepackage{multirow}
\usepackage{tikz}
\usepackage{xstring}
\usepackage{makecell}
\usepackage{grffile}
\usepackage{xr}
\usepackage{bbm}
\usepackage{ifthen}
\usetikzlibrary{decorations.markings}
\usetikzlibrary{positioning}
\usetikzlibrary{arrows.meta}
\usepackage{tikz-feynman}
\usetikzlibrary{calc}
\tikzfeynmanset{warn luatex=false}

\makeatletter
\def\maketag@@@#1{\hbox{\m@th\normalfont\normalsize#1}}
\makeatother

\allowdisplaybreaks[4]

\makeatletter     
\renewcommand\onecolumngrid{
\do@columngrid{one}{\@ne}%
\def\set@footnotewidth{\onecolumngrid}
\def\footnoterule{\kern-6pt\hrule width 1.5in\kern6pt}%
}

\makeatletter
\AtBeginDocument{\let\LS@rot\@undefined}
\makeatother

\usepackage{times}
\usepackage{graphicx}
\usepackage{float}
\usepackage{latexsym,amsmath,amssymb,bm,euscript}
\usepackage{color}
\usepackage{epstopdf}
\usepackage[colorlinks=true,linkcolor=blue,citecolor=blue,urlcolor=blue]{hyperref}
\usepackage{cleveref}

\usepackage{ulem}
\usepackage{graphicx}
\usepackage{verbatim}
\usepackage{physics}

\usepackage{lineno}
\usepackage{bm}
\usepackage{mathptmx}
\usepackage{mathrsfs}
\usepackage[marginal]{footmisc}
\usepackage{tikz}
\usetikzlibrary{decorations.pathreplacing}

\usepackage{ytableau}

\newcommand{\be}[0]{\begin{equation}}
\newcommand{\ee}[0]{\end{equation}}
\def\ba#1\ea{\begin{align}#1\end{align}}  
\def\baa#1\eaa{\begin{align}#1\end{align}}

\crefname{appendix}{Appendix}{Appendices}
\crefname{equation}{Eq.}{Eqs.}
\crefname{figure}{Fig.}{Figs.}
\crefname{table}{Table}{Tables}
\crefname{section}{Section}{Sections}
\crefname{enumi}{Point}{Points}
\creflabelformat{appendix}{[#2#1#3]}
\crefmultiformat{figure}{Figs.~#2#1\xdef\mycreffirstarg{#1}#3}%
 { and~#2#1#3}{, #2#1#3}{ and~#2#1#3}

\makeatletter

\begin{document}

\date{\today}

\title{Residual entropy from temperature incremental Monte Carlo method}
\author{Zenan Dai}
\affiliation{Key Laboratory of Artificial Structures and Quantum Control (Ministry of Education), School of Physics and Astronomy, Shanghai Jiao Tong University, Shanghai 200240, China}
\affiliation{Zhiyuan College, Shanghai Jiao Tong University, Shanghai 200240, China}
\author{Xiao Yan Xu}
\email{xiaoyanxu@sjtu.edu.cn}
\affiliation{Key Laboratory of Artificial Structures and Quantum Control (Ministry of Education), School of Physics and Astronomy, Shanghai Jiao Tong University, Shanghai 200240, China}
\affiliation{Hefei National Laboratory, Hefei 230088, China}
\date{\today}

\begin{abstract}  
Residual entropy, which reflects the degrees of freedom in a system at absolute zero temperature, is crucial for understanding quantum and classical ground states. Despite its key role in explaining low-temperature phenomena and ground state degeneracy, accurately measuring residual entropy remains a difficult task owing to computational limitations. In this Letter, we introduce the temperature incremental Monte Carlo (TIMC) method, our approach to overcoming these challenges. The TIMC method incrementally calculates the partition function ratio of neighboring temperatures within Monte Carlo simulations, enabling precise entropy calculations and revealing other temperature-dependent properties in a single computational sweep of temperatures. We have rogorously tested TIMC on several complex systems, including the frustrated antiferromagnetic Ising model on both C60 and 2D triangular lattices, the Newman-Moore glassy model, and a 2D quantum transverse field Ising model. Notably, our method overcomes the difficulties encountered in partition function measurements when mapping $d$-dimensional quantum models to $d+1$-dimensional classical counterparts. These challenges arise from singular interactions that emerge in the small $\Delta_\tau$ limit during the quantum-to-classical mapping procedure. The TIMC method enables precise entropy calculations across the entire temperature range, as demonstrated in our studies of frustrated spin models, glassy phases, and phases exhibiting spontaneous symmetry breaking. This method's capability to calculate residual entropy could provide insights when applied to systems lacking analytical solutions.
\end{abstract}

\maketitle 

{\it Introduction}. The study of residual entropy has a storied history, tracing back to the pioneering analysis of water ice, a quintessential example that first highlighted the concept~\cite{pauling1935structure,Lieb1967}. In contemporary physics, spin ice represents a notable instance of residual entropy. Similar forms of geometrically frustrated magnetism have garnered interest for their potential applications in computation, data storage and refrigeration technologies~\cite{kitaev2003fault,kitaev2006anyons,Caravelli2020,heyderman2022spin,balents2010spin,xiang2024giant}. The significance of residual entropy extends to its relationship with ground state degeneracy, which plays a pivotal role in the classification of phases of matter including phases exhibiting spontaneous symmetry breaking, topological orders~\cite{wen1990topological} and fracton phases~\cite{nandkishore2019fractons,pretko2020fracton}. Despite its importance, the analytical calculation of residual entropy remains a formidable challenge, with only a handful of models yielding analytical exact solutions~\cite{Lieb1967,moessner2001ising,nandkishore2019fractons,pretko2020fracton,march2016exactly}.

Conversely, the Monte Carlo method stands as a vital numerical tool, renowned for its unbiased approach to complex physical systems. Traditional Monte Carlo techniques, however, encounter limitations when measuring entropy. While widely used, existing methods face significant challenges: the Wang-Landau method~\cite{wangEfficientMultipleRange2001} suffers from slow convergence and low efficiency, and thermodynamic integration methods~\cite{binder1985monte,zukovicResidualEntropySpins2013a} introduce systematic errors through their numerical integration procedures. Although direct computation of the partition function offers a pathway to determine entropy, it is plagued by poor distribution as an observable. However, breakthroughs in incremental calculation and reweighting approaches~\cite{Ferrenberg1989,Jarzynski1997,neal2001annealed,Alba2017}, which have proven successful in computing challenging quantities like entanglement entropy, suggest a promising new direction for accurately determining residual entropy~\cite{Emidio2020,demidio2022universal,ArXiv2023Liaoa,ArXiv2023Zhangd,zhou2024incremental}.

To estimate residual entropy accurately, one must compute the entropy at various temperatures and extrapolate these values to absolute zero. Previous incremental methods required executing a complete incremental procedure to obtain the partition function at each temperature~\cite{Emidio2020,demidio2022universal,ArXiv2023Liaoa,ArXiv2023Zhangd,zhou2024incremental}, a process often seen as laborious. Addressing this, we introduce the temperature incremental Monte Carlo (TIMC) method in this Letter. Our method assigns each incremental step with a specific temperature, thereby allowing the acquisition of the partition function across the entire temperature range within a single simulation. Our TIMC method offers two key advantages: it achieves significantly better efficiency than the Wang-Landau algorithm, and while matching the computational speed of thermodynamic integration, it eliminates the systematic errors that typically plague integration-based approaches. These improvements make TIMC both more practical and more reliable for calculating residual entropy.

We thoroughly tested TIMC's capabilities through comprehensive calculations of residual entropy across many systems, encompassing the frustrated classical antiferromagnetic Ising model on both C60~\cite{samuelSolutionIsingModel1993,rasettiDimerCoveringIsing1981,krotoC60Buckminsterfullerene1985} and triangular lattices, the Newman-Moore glassy model~\cite{Spinglass}, and the quantum Ising model both on a square lattice and on a trianglular lattice. The results are convincing: TIMC not only delivers superior accuracy compared to conventional statistical methods but also provides a robust framework for free energy calculations that remains stable even as the imaginary time discretization approaches zero. A key advantage of TIMC over existing incremental methods is its ability to compute entropy values across multiple temperatures in a single computational run, making it both efficient and practical. Looking ahead, TIMC shows great promise for applications in fermionic and bosonic systems, offering a powerful tool for exploring and identifying previously unknown phases of matter and their transitions.

{\it Model and Methods}. Now we introduce the general formula of the TIMC method. The entropy $S$, internal energy $U$, free energy $F$ and partition function $Z$ at inverse temperature $\beta=\frac{1}{T}$ are related by $S[\beta]=\ln Z[\beta]+\beta U[\beta]$ and $F[\beta]=-\frac{1}{\beta}\ln Z[\beta]$. It is usually easy to evaluate internal energy $U[\beta]$; therefore, to get entropy $S[\beta]$, it is crucial to calculate the partition function $Z[\beta]$. In the TIMC method, the partition function $Z[\beta]$ is calculated as a series product of the partition function ratio of neighboring temperatures
\begin{equation}
Z[\beta]= Z[0]\prod_{k=0}^{M-1} \frac{Z_{k+1}}{Z_k},
\end{equation}
where $M$ is the number of incremental steps, which should be set proportional to $\beta$ and system size to make sure each ratio is at $\mathcal{O}(1)$ when increasing $\beta$ and system size. It is worth noting that the small $M$ limit corresponds to the histogram reweighting method~\cite{Ferrenberg1989}, which extrapolates thermodynamic properties at neighboring parameters using configurations sampled at a single point. However, this approach becomes ineffective for large systems owing to exponentially decreasing overlap between energy distributions. $Z_k = Z\left[ \frac{k}{M}\beta \right]$ is the partition function at inverse temperature $\frac{k}{M}\beta$, which is the \textit{crucial difference} comparing to the previous incremental method~\cite{Emidio2020,demidio2022universal,ArXiv2023Liaoa,ArXiv2023Zhangd,zhou2024incremental} where the intermediate $Z_{k=1,\cdots,M-1}$ is not related to partition function of any real physical system. We also note that $Z[0]=Z_{k=0}$ is the partition function at infinite high temperature, which equals the total number of degrees of freedom of the system.

 In the TIMC method, each incremental step involves computing a ratio of partition functions at successive inverse temperatures. Specifically, we calculate the ratio:

\begin{equation}
\frac{Z_{k+1}}{Z_k} = \frac{\sum_{s}w_{s,k}O_{s,k}}{\sum_s w_{s,k}}
\end{equation}
Here, $O_{s,k}$ represents the ratio of Boltzmann weights for a given configuration $s$ between two successive steps:

\begin{equation}
O_{s,k}=\frac{w_{s,k+1}}{w_{s,k}}.
\end{equation}

The Boltzmann weight $w_{s,k}$ corresponds to the configuration $s$ at an inverse temperature of $\frac{k}{M}\beta$. Since $Z_k$ denotes the partition function at this inverse temperature, we can calculate the partition function at any desired inverse temperature within our range using these ratios. This ability to determine partition functions across a spectrum of temperatures is a key advantage of the TIMC method. Additionally, as the Boltzmann weights are readily available from this process, we can simultaneously compute other physical observables at each inverse temperature $\frac{k}{M}\beta$. This dual capability is particularly advantageous compared with other incremental methods, where intermediate steps typically lack physical interpretation. TIMC ensures that calculations at each intermediate temperature provide physically meaningful results while maintaining the efficiency of the incremental approach.

For the classical model, one can write the Boltzmann weight as $w_{s,k} = e^{-E_{s,k}^{\text{eff}}}$, where the effective energy $E_{s,k}^{\text{eff}} = \frac{k}{M} E_s$, such that $O_{s,k}$ has a very simple form, $O_{s,k}=e^{-E_s /M}$, with $E_s$ the dimensionless energy ($\beta$ is absorbed into the energy) of configuration $s$. 

For the quantum model, if the quantum to classic mapping allows writing the Boltzmann weight $w_{s,k}$ as the form $e^{-E_{s,k}^{\text{eff}}}$, then $O_{s,k}$ has similar formula as the classical case. We take the transverse field Ising model on a square lattice as an example.
The Hamiltonian is
\begin{equation}
H = -J \sum_{\langle i,j \rangle }\hat{Z}_{i} \hat{Z}_{j} - h \sum_{i}^{} \hat{X}_{i}
\label{quantum Ising}
\end{equation}
where $h$ is the strength of the traverse field for the $x$-direction spin $\hat{X}$, $J$ is the nearest neighbor coupling for the $z$-direction spin $\hat{Z}$. We set the lattice size to be $L\times L$.

First, we map the partition function of the quantum Ising model to the partition function of a higher dimensional classical Ising model.
\begin{equation}
Z_k = \sum_s e^{-E_{s,k}^{\text{eff}}} \equiv \sum_s w_{s,k}
\end{equation}
where the effective energy has the form
\begin{equation}
E_{s,k}^\text{eff} =- \frac{k}{M} \Delta_\tau J\sum_{\langle i,j\rangle,l} s_{i,l} s_{j,l} - \gamma_k \sum_{i,l} \left( s_{i,l} s_{i,l+1} - 1\right)-V\ln \Lambda_k,
\end{equation}
with $V=L^2L_\tau$, $\Lambda_k = \cosh \left( \frac{k}{M} \Delta_\tau h \right)$, $\gamma_k = -\frac{1}{2}\ln \tanh \left( \frac{k}{M} \Delta_\tau h \right) $. It follows that $O_{s,k}$ has a simple form
$O_{s,k} = \exp\left({-E_{s,k+1}^{\text{eff}}+E_{s,k}^{\text{eff}}}\right)$. It is notable out that the $\gamma_k$ term diverges at $k = 0$. However, as in this case, the nearest neighbor bonding in the imaginary time direction is infinite, such that we always have $s_{i,l} s_{i,l+1} -1 =0$, which makes the contribution of $\gamma_k$ term in the effective energy $E_{s,k}^{\text{eff}}$ vanish. Therefore $Z[0]=Z_{k=0}$ is still well defined. We further note that although $\gamma_k$ is singular in the limit of $\Delta_\tau \rightarrow 0$, the difference $\gamma_{k+1} - \gamma_{k}$ in the $\Delta_\tau \rightarrow 0$ is well defined,
\begin{equation}
\lim_{\Delta \tau \to 0} ( \gamma_{k+1}-\gamma_k )= \frac{1}{2 k}.
\end{equation}
This property ensures the stability of $O_{s,k}$ calculations, and thus the partition function, even in the limit of small $\Delta_\tau$.

In instances where the quantum-to-classical mapping does not simplify the Boltzmann weight, computing of $O_{s,k}$—the observable of interest at incremental step $k$—may require additional effort. Nevertheless, this challenge is manageable. For instance, within the framework of the TIMC method applied to determinant Quantum Monte Carlo (DQMC)~\cite{Blankenbecler1981qmcc,Assaad2008compumanybody,loh1992stable}, the evaluation of $O_{s,k}$ can be efficiently executed using fast or delayed update algorithms~\cite{alvarez2008new,nukala2009fast,gull2011submatrix,sun2023delay}. These techniques streamline the update process within the Monte Carlo simulation, minimizing computational overhead.

In the most demanding scenarios, where $O_{s,k}$ remains difficult to compute through straightforward methods, one might resort to parallel tempering strategies. This approach involves running multiple simulations at different temperatures in parallel, which can facilitate the direct calculation of the ratio $O_{s,k}=\frac{w_{s,k+1}}{w_{s,k}}$. While parallel tempering is a general technique available to Monte Carlo methods, its integration with TIMC is particularly useful as it ensures reliable sampling of Boltzmann weight ratios between neighboring temperatures.

\begin{figure}[t]
    \centering
    \includegraphics[width=3.5in]{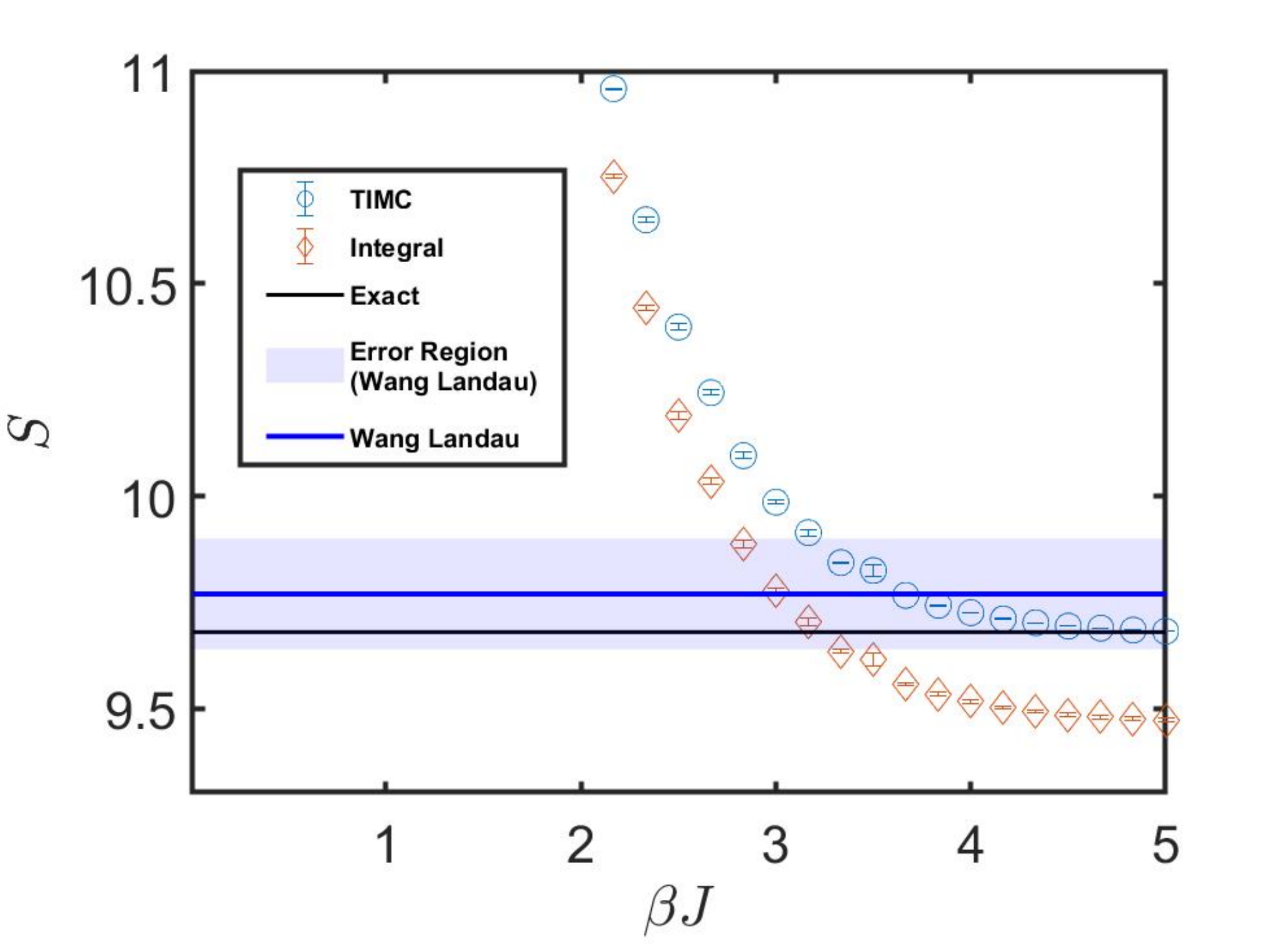}
    \caption{Entropy as a function of inverse temperature for antiferromagnetic Ising model on the C60 lattice. The Wang-Landau and the exact results are at zero temperature. $M=30$ is used here in TIMC.}
    \label{FG1}
\end{figure}

{\it Antiferromagnetic Ising model}.---
To demonstrate the TIMC method's precision, we applied it to the antiferromagnetic Ising model on both C60 and triangular lattices. The Hamiltonian is given by $H= J \sum_{\langle i,j \rangle} \hat{Z}_i\hat{Z}_j$, with $J>0$ indicating antiferromagnetic coupling and consideration given solely to nearest-neighbor interactions. We begin with the C60 lattice, composed of 12 pentagons and 20 hexagons. Frustration within the pentagons leads to a substantial ground state degeneracy for this model on the C60 lattice~\cite{samuelSolutionIsingModel1993,rasettiDimerCoveringIsing1981}, quantified as 16,000 distinct states—a figure that can be established via combinatorial analysis. The entropy temperature function derived from our TIMC method, as illustrated in Fig.\ref{FG1}, is in excellent agreement with the analytical expectation that predicts a zero-temperature entropy of $\ln(16000)\approx 9.6803$~\cite{suppl}. Our numerical results yield an entropy value of $S = 9.6824\pm0.0004$ at $\beta/J=5$, demonstrating remarkable consistency with theoretical predictions. Notably, the entropy value at $\beta/J=5$ closely approximates the residual entropy at zero temperature, attributable to a sizable excitation gap on the order of $J$. Note that for the model without a sizable excitation gap, one should go to lower temperatures to extrapolate to zero temperature.

In addition, we compared the results obtained using our method with those derived from the integral method~\cite{zukovicResidualEntropySpins2013a} employing the same number of increments, as well as the Wang-Landau method~\cite{wangEfficientMultipleRange2001}, which utilized comparable scales of standard Metropolis–Hastings updates, as illustrated in Fig.~\ref{FG1}. The comparison indicates that our TIMC method achieves greater accuracy than the Wang-Landau method and effectively avoids the systematic errors associated with the integral method.

\begin{figure}[b]
    \centering
    \includegraphics[width=3.0in]{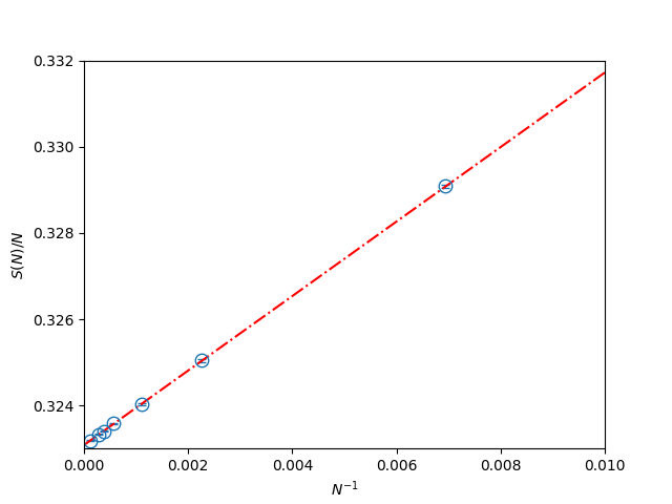}
    \caption{Entropy as a function of the lattice size for the antiferromagnetic Ising model on the triangular lattice. Dashed line: linear fit; dots: numerical results.}
    \label{FG2}
\end{figure}

The triangular lattice serves as another prototypical example of geometric frustration. For the antiferromagnetic Ising model on this lattice, the degree of ground state degeneracy is significantly higher. According to analytical exact solutions~\cite{PhysRev.79.357} and high-precision numeric results~\cite{zukovicResidualEntropySpins2013a,Vanderstraeten2018}, this degeneracy grows exponentially with the increase in system size, implying that entropy should similarly exhibit growth as a function of system size. The analytical calculation for the entropy per site is $ \frac{3}{\pi} \int_0^{\frac{\pi}{6}}\ln(2 \cos(\omega)) d \omega \approx 0.323066$.
Our TIMC method results, presented in Fig.~\ref{FG2}, confirm this behavior: entropy indeed increases with the system size $N$. The derived slope from our data yields an entropy per site value of $0.323080(7)$, which is in close alignment with the exact analytical results.


\begin{figure}[t]
    \centering
    \includegraphics[width=3.5in]{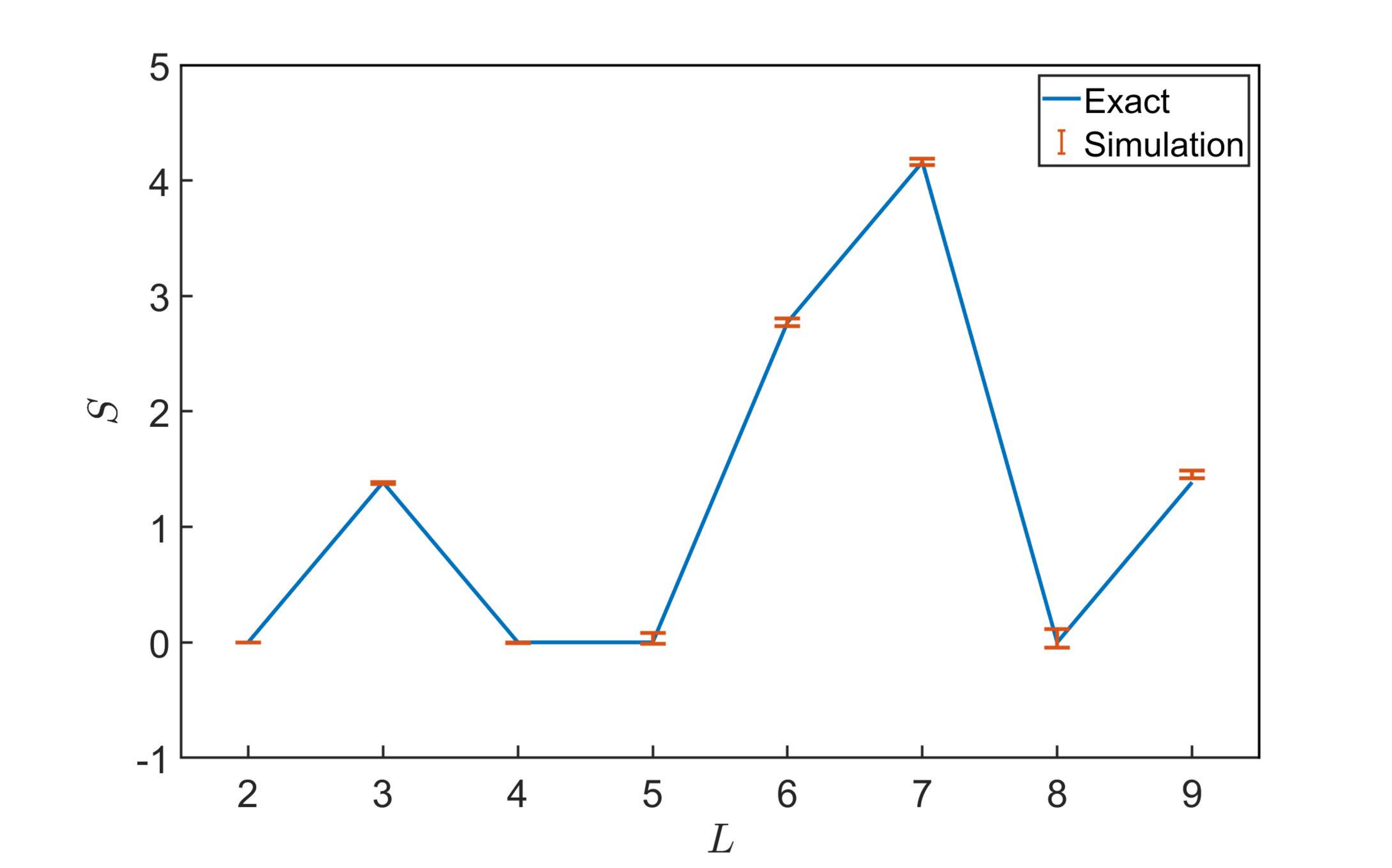}
    \caption{Entropy as a function of the lattice size for the Newman-Moore glassy model on the triangular lattice with $ L_x=L_y=L$ and periodic boundary conditions. Solid line: analytical solutions; dots: numerical results.}
    \label{FG3}
\end{figure}

{\it The Newman-Moore glassy model}. The TIMC method's versatility extends to the computation of residual entropy for glassy models. For illustrative purposes, we focus on the Newman-Moore glassy model~\cite{Spinglass}. The model's Hamiltonian is expressed as
\begin{equation}
H=\frac{1}{2} J \sum_{i, j, k \text { in } \bigtriangledown} \sigma_i \sigma_j \sigma_k,
\end{equation}a
where each $\sigma_i$ takes a value of $\pm 1$. The summation is restricted to the interactions among three spins forming an inverted triangle on the triangular lattice. The glassiness of this model is attributed to kinetic constraints\cite{garrahan2002glassiness}. The ground state degeneracy can be determined by counting periodic orbits of a cellular automaton and has number-theoretic properties~\cite{martin1984algebraic} and boundary condition dependence \cite{sfairopoulos2023boundary}. Here, we consider only the system with $L_x=L_y=L$ and periodic boundary conditions. Interestingly, the degeneracy does not follow a monotonic trend; rather, it oscillates with the system size. It has been established~\cite{Spinglass,sfairopoulos2023boundary} that the model possesses a unique ground state when the system size is a power of two ($2^k, k\in \text{Integer}$).

As depicted in Fig.\ref{FG3}, our TIMC method's computation of residual entropy across various system sizes agrees well with analytical predictions, despite the complex nature of glassy systems. 
This underscores the TIMC method's robustness and its potential as a reliable tool for probing the intricate landscapes of glassy systems.


{\it Quantum Ising model}. The TIMC method is equally adept at addressing quantum models, exemplified by our application to the quantum Ising model. We first consider the ferromagnetic transverse field Ising model on a square lattice, as described by the Hamiltonian in Eq.~\eqref{quantum Ising}. At zero temperature, this system undergoes a quantum phase transition from a ferromagnetic phase to a paramagnetic phase around the critical point $h/J\approx 3.04$. In the ferromagnetic phase, characterized by spontaneous $Z_2$-symmetry breaking, the ground state degeneracy is twofold. Conversely, in the paramagnetic phase, the transverse field aligns the spins in the $x$-direction, resulting in a non-degenerate ground state.

\begin{figure}[h]
    \centering
\includegraphics[width=0.5\textwidth]{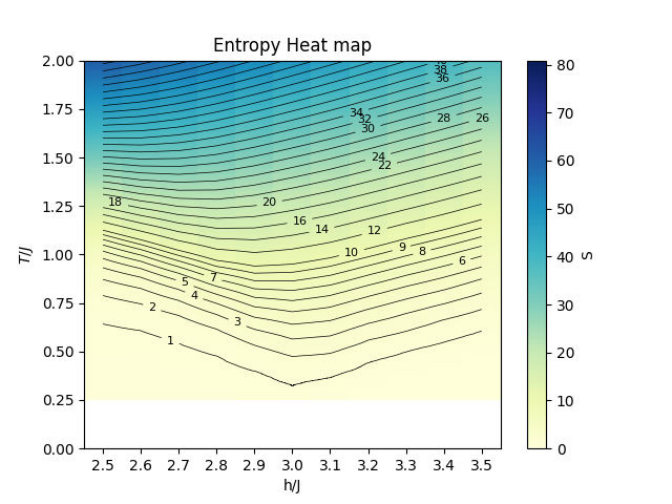}
\caption{Entropy landscape of transverse field Ising model on a square lattice. The smoothed entropy data $S(T)$ at $ h = 2.5,2.6,2.7,2.8,2.9,3.0,3.1,3.2,3.3,3.4,3.5 $ for lattice size $ L = 16 $ are presented.}
    \label{FG6}
\end{figure}

Achieving precise entropy measurements near critical points presents significant challenges. This is particularly evident when attempting to verify the exact zero-entropy at $h>h_c$ and $\ln 2$ entropy at $h<h_c$, as the calculation involves $S[\beta] = \ln Z[\beta] + \beta U[\beta]$, where both terms are extensive quantities that largely cancel each other, requiring exceptional precision in $\ln Z[\beta]$ measurements. Our detailed analysis, presented in Fig.~\ref{FG6}, maps the entropy landscape of the quantum Ising model near the phase transition point. At low temperatures, we clearly identify two distinct regimes: a plateau at approximately $\ln(2)$ for $\frac{h}{J}<3.04$ and a plateau at $0$ for $\frac{h}{J}>3.04$. Notably, we observe pronounced entropy peaks around the transition points. The iso-entropy contours we obtained demonstrate a significant adiabatic cooling effect near the quantum critical point (QCP) - a phenomenon we also identified near the QCP of the transverse field Ising model with antiferromagnetic interactions on a triangle lattice~\cite{suppl}.

{\it Conclusion and outlook}.---
In conclusion, our study has successfully demonstrated the robustness and versatility of the Temperature Incremental Monte Carlo (TIMC) method across a range of models, both classical and quantum. The TIMC method has proven itself as a powerful computational tool, capable of accurately calculating residual entropies in systems with varying degrees of ground state degeneracy.

For classical models, such as the antiferromagnetic Ising model on C60 and triangular lattices, the TIMC method not only confirmed the expected exponential increase in ground state degeneracy with system size for the triangular lattices antiferromagnetic Ising model but also provided entropy values in excellent agreement with analytical predictions. Similarly, in the context of glassy models like the Newman-Moore model, the TIMC method adeptly navigated the oscillatory degeneracy behavior, offering results consistent with complex mathematical solutions. For systems where exact transfer matrix solutions are unavailable, TIMC's polynomial scaling with system size offers a computational advantage over transfer matrix methods, which typically scale exponentially. However, for systems with exact transfer matrix solutions, those methods remain preferable.

In the quantum domain, our application of the TIMC method to the transverse field Ising model demonstrates its capability to precisely compute entropy and determine ground state degeneracy. Through entropy measurements, TIMC can clearly distinguish between the unique ground state of the paramagnetic phase and the twofold degeneracy of the ferromagnetic phase at zero temperature. Most notably, our analysis of the iso-entropy contours reveals a significant adiabatic cooling effect, highlighting TIMC's capability to capture subtle quantum thermodynamic phenomena.

Looking forward, the TIMC method holds promise for exploring more intricate classical and quantum systems, and could be adapted to tackle both fermionic and bosonic models. For example, it would be interesting to apply it to higher-dimensional fully or partially frustrated models~\cite{moessner2001ising,Netz1991,Jalabert1991}. Further more, for the model with a sign problem, the TIMC method can map the entropy landscape at high temperature where the sign problem is less severe, and thus may help identify possible phases at lower temperatures, for example, if there is any tendency to form superconductivity, the entropy should drop significantly. 

The TIMC method's success in handling extensive degeneracies and capturing critical phenomena at phase transitions opens up avenues for investigating many-body systems. The potential applications of this method span a multitude of fields, from condensed matter physics to quantum information, where understanding entropy and ground state degeneracy is crucial.

\textit{Note added}. Recently, an independent study by Ding \textit{et al.}~\cite{ding2024} appeared, proposing a similar physical parameter-based incremental approach for computing thermodynamic quantities. Their method shares conceptual similarities with our TIMC framework, providing complementary validation of the effectiveness of temperature-incremental strategies in statistical physics calculations. Furthermore, as brought to our attention by the referee, similar concepts were previously explored by Pollet \textit{et al.}~\cite{polletTemperatureChangesWhen2008} in earlier work.

\begin{acknowledgments}
{\it Acknowledgements}.---We thank Tarun Grover for helpful discussions, and Konstantinos Sfairopoulos for helpful comments on the draft. This work is supported by the National Key R\&D Program of China (Grant No. 2022YFA1402702, No. 2021YFA1401400), the National Natural Science Foundation of China (Grants No. 12274289, No. 12447103), the Innovation Program for Quantum Science and Technology (under Grant no. 2021ZD0301902), Yangyang Development Fund, and startup funds from SJTU.
The computations in this paper were run on the Siyuan-1 and $\pi$ 2.0 clusters supported by the Center for High Performance Computing at Shanghai Jiao Tong University.
\end{acknowledgments}

\appendix
\setcounter{equation}{0}
\setcounter{figure}{0}
\setcounter{table}{0}
\setcounter{page}{1}
\renewcommand{\thetable}{S\arabic{table}}
\renewcommand{\theequation}{S\arabic{equation}}
\renewcommand{\thefigure}{S\arabic{figure}}
\renewcommand{\bibnumfmt}[1]{[S#1]}
\renewcommand{\citenumfont}[1]{S#1}
\setcounter{secnumdepth}{3}

\section{Detail in Numerical Monte Carlo}

\subsection{Gound state number of anti-ferromagnetic Ising model on $C_{60}$}
In this section, we establish that the ground state of the system contains 16,000 configurations for anti-ferromagnetic Ising model on $ C_{60} $ molecule. For clarity in description, we define an edge as {\it{consistent}} if the spins at its two endpoints are antiparallel, and, {\it inconsistent} if the spins are parallel.

\begin{figure}[H]
    \centering
    \subfloat{\includegraphics[width=0.3\textwidth]{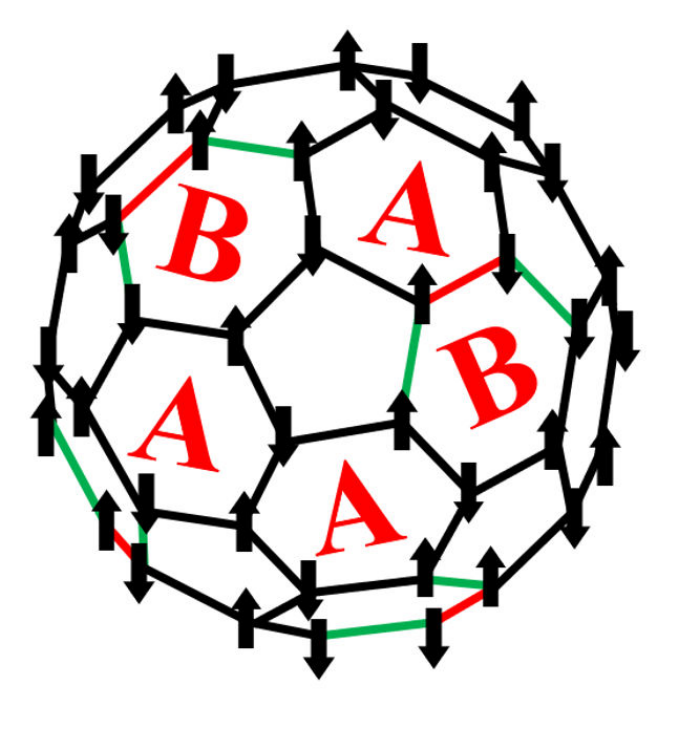}}
    \caption{Example of Spin configuration that minimizes the energy.}
    \label{fig:s1}
\end{figure}

First, we analyze the configuration that minimizes the energy on each pentagon. For each pentagon, a inconsistent edge is identified and marked in green, as shown in \cref{fig:s1}. In total, the system contains 12 inconsistent edges.

Having minimized the energy on each pentagon, we now focus on the position of each inconsistent edges to further minimize the global energy. We find that all remaining edges can be consistent, provided that each hexagon has either two or zero inconsistent edges. Consequently, the problem of determining the ground states is transformed into counting the number of valid inconsistent edge configurations. This transformation represents a two-to-one mapping, as flipping all the spins does not alter the configuration of inconsistent edges.

We classify the hexagons into two types: $A$ type and $B$ type. The $ A $ type hexagon has zero inconsistent, while the $ B $ type hexagon has two inconsistent edges as shown in \cref{fig:s1}. Additionally, we introduce a new special red line in \cref{fig:s1}, which connects two inconsistent green edges. We find that a distinct inconsistent edge configuration can be obtained by swapping $ A $, $ B $ type hexagon on either side of the red line. Therefore, there exists a $ 2^{6} $ to 1 mapping between the configuration of the red lines and the configuration of the green lines.

\begin{figure}[H]
    \centering
    \subfloat{\includegraphics[width=0.3\textwidth]{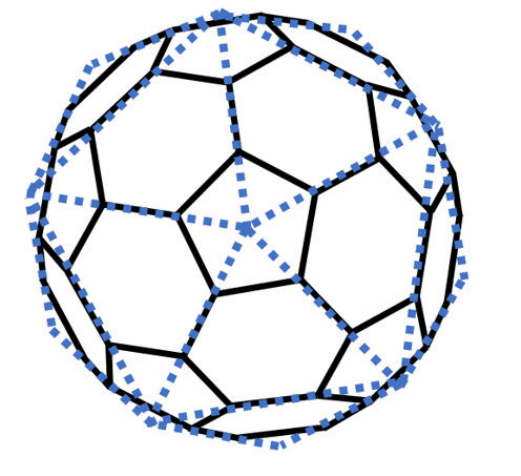}}
    \caption{Possible red line positions in \cref{fig:s1}, denoted as blue dotted lines here.}
    \label{fig:s2}
\end{figure}

\begin{figure}[H]
    \centering
    \subfloat{\includegraphics[width=0.3\textwidth]{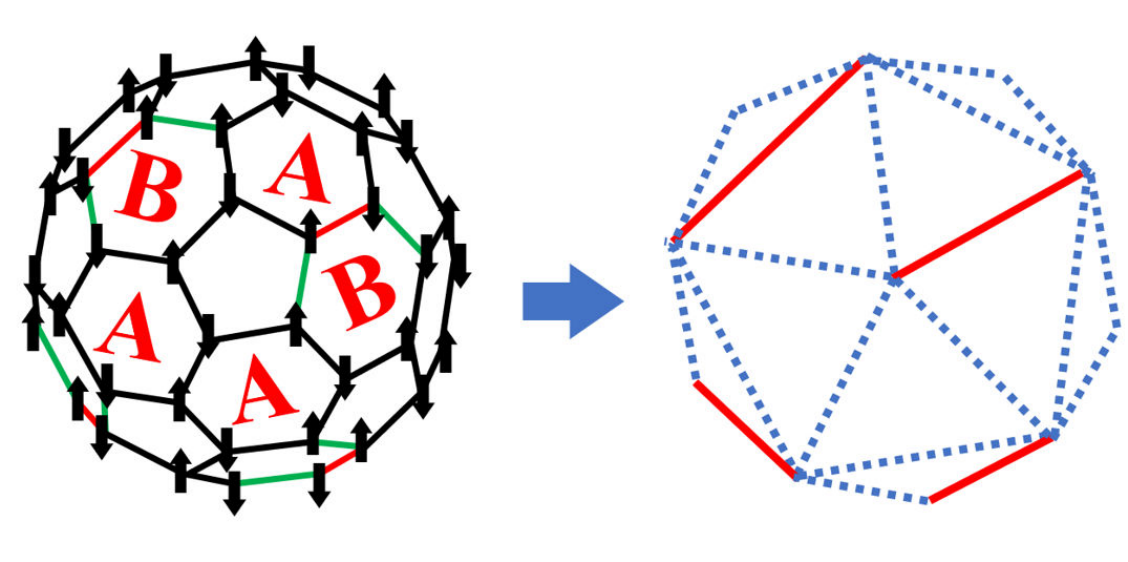}}
    \caption{Mapping between spin configurations and red line configurations.}
    \label{fig:s3}
\end{figure}

Now, we count the possible configurations of the red lines. All the red lines together form a regular icosahedron, as illustrated in \cref{fig:s2}, and the mapping becomes clearer in \cref{fig:s3}. For the red line configuration on the regular icosahedron, the following rule applies: at each vertex, there is exactly one red line connected to it. This follows from the fact that each pentagon contains only one inconsistent edge.

The number of valid red line configurations can be determined as follows. First, we consider the top and bottom vertices, for which there are $ 5 \times 5 $ choices to connect them. Next, we connect the remaining 8 vertices in the middle using 4 additional red lines. After enumerating the possibilities, we find that there are 5 ways to make these connections. Therefore, the total number of red line configurations is $ 5\times 5 \times 5 = 125 $, leading to $ 16,000 $ distinct ground states.


\subsection{Numerical calculation of anti-ferromagnetic Ising model on $C_{60}$}

In numerical realization, we use the standard Metropolis algorithm to update the spin configuration. We take $ T= 0.2J $ as the lowest temperature, and perform 30 increments ($M=30$). we average over the result of 100 bins.  In each bin, we initialize a random spin configuration and iterate until thermalization. After thermalization, we apply the Metropolis update $ 10^6 $ times. In total, we perform $ 3 \times 10^9 $ Metropolis updates.

\subsection{Numerical calculation of anti-ferromagnetic Ising model on triangular lattice}
\label{sec:g1d}

In the numerical implementation, we utilize the standard Metropolis algorithm in combination with the geometric cluster algorithm and the Wolff algorithm to update the spin configuration. We take $ T= 0.2J $ as the lowest temperature and perform 320 increments ($M=320$), for each increment, a random spin configuration is initialized and iterated until thermalization. After thermalization, we execute the combined update $ 10^6 $ times. In total, $ 3.2 \times 10^8 $ combined updates are performed. Using linear regression, we estimate the residual entropy of the model in the thermodynamic limit to be $ 0.323080(7) $, which is consistent with the theoretical value $ 0.3230659669 $.

\subsection{Numerical calculation of Newman-Moore Glassy Model}
The Hamiltonian of the Newman-Moore glassy model \cite{Spinglass} is given by

\begin{equation}
  H=\frac{1}{2} J \sum_{i, j, k \text { in } \bigtriangledown} \sigma_i \sigma_j \sigma_k,
\end{equation}

\begin{figure}[H]
    \centering
    \subfloat{\includegraphics[width=0.4\textwidth]{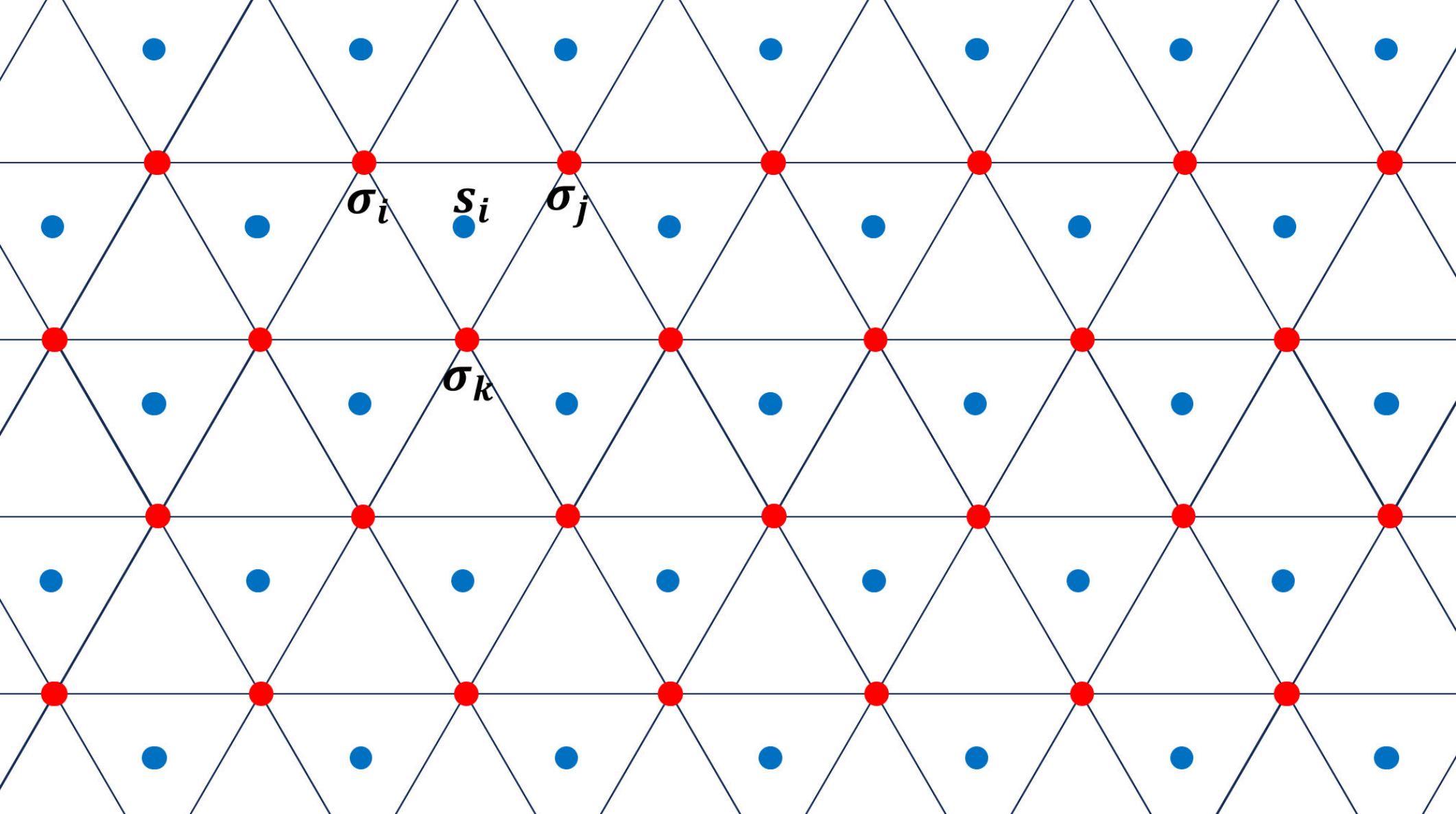}}
    \caption{Mapping between the $ \sigma_i $ (red) to $ s_j $ (blue).}
    \label{fig:maps}
\end{figure} 

In our numerical implementation, we transform the interacting Ising model with unrestricted updates into a non-interacting spin model with constrained updates. Specifically, the product of the three spins at the vertices of a triangle $\sigma_i \sigma_j \sigma_k$ is mapped to a single spin on the center of the triangle, denoted as $ s_i $ as depicted in \cref{fig:maps}. This transformation results in a non-interacting spin system.

\begin{equation}
  H=\frac{1}{2} J \sum_{i } s_i
\end{equation}

For the original model, we can flip the spin $ \sigma_{i} $ freely, but in the mapped model, we must flip the three spins $ s_i $ on the vertex of an $ \triangle $ simultaneously which corresponds to the flip of a single spin $ \sigma_i $ .

In numerical practice, we begin with a random spin configuration of the Newman-Moore glassy model and map it to the non-interacting model. The entropy at $ T = 0.1 J$ is taken as an approximation of the residual entropy, which proves to be sufficiently accurate given the small variance. The temperature range is divided into $ 20 $ increments ($M=20$), for each increment, we sample 100 bins, performing $ 2\times 10^6 $ Metropolis update in each bin. Consequently, a total of $ 4 \times 10^9 $ Metropolis updates are conducted for each lattice size.

\section{Comparison with Wang-Landau Method}

In this section, we compare the Wang-Landau method\cite{wangEfficientMultipleRangeRandom2001} and the TIMC method in the context of calculating the residual entropy.  To ensure a fair comparison, we terminate the Wang-Landau method after the same number of Metropolis updates and evaluate the accuracy of the residual entropy obtained by each approach.

For Wang-Landau method for anti-ferromagnetic on $C_{60}$ lattice, we divide the energy into 60 windows. To extract the residual entropy, the lowest energy window includes only the ground state energy. Initially, a stabilized density of states distribution is obtained using the conventional Wang-Landau method. The flatness condition for the sampled histogram requires that the counts in each energy window differ by no more than $ 1\% $. The update step for the density of states begins at $ 5 \times 10^{-3} $ and is gradually reduced whenever the sampled histogram becomes flat. The process continues until the histogram is flat at an update step of $ 10^{-4} $ . 

After stabilization, the Wang-Landau method is continued to sample the entropy every $ 10^6 $ updates, yielding a total of 200 sampled points. Over the entire procedure, $ 3.26 \times 10^9 $ Metropolis updates are performed. The resulting residual entropy is $ 9.77 \pm 0.13$, which is less accurate compared to the TIMC method as depicted in \cref{FG1}.


This outcome is expected, as the conventional Wang-Landau method is designed to obtain the entire density of states distribution rather than specifically targeting the residual entropy.

\section{Comparison with Integral Method}

In this section, we compared the diffrence between the integral method and TIMC algorithm, taking anti-ferromagnetic ising model on $ C_{60} $ as an example. In our consideration, we denote the integral method as integration as follows \cite{nealAnnealedImportanceSampling1998}

\begin{equation}
\begin{aligned}
  S(\beta) = S(0)-\int_0^{\beta} E(\beta) d\beta
\end{aligned}
\label{}
\end{equation}

\hspace{1cm}

In general, for small enough increment $ \Delta \beta \to d \beta $, the integration error would be negligible. But the practical calculation uses finite $\Delta \beta$ which introduce a systematic error. Instead, our TIMC method is unbiased. As shown in \cref{FG1}, the value of integral method deviates from the exact results due to the systematic error, while the TIMC results match the exact results well. Note that $\Delta \beta = \frac{1}{6J} $ here and $M=30$.

\section{Quantum Ising model Calculation}
\subsection{Square Lattice Quantum Ising model}
\clearpage
\begin{widetext}

\begin{figure}[H]
    \centering
    \subfloat{\includegraphics[width=\textwidth]{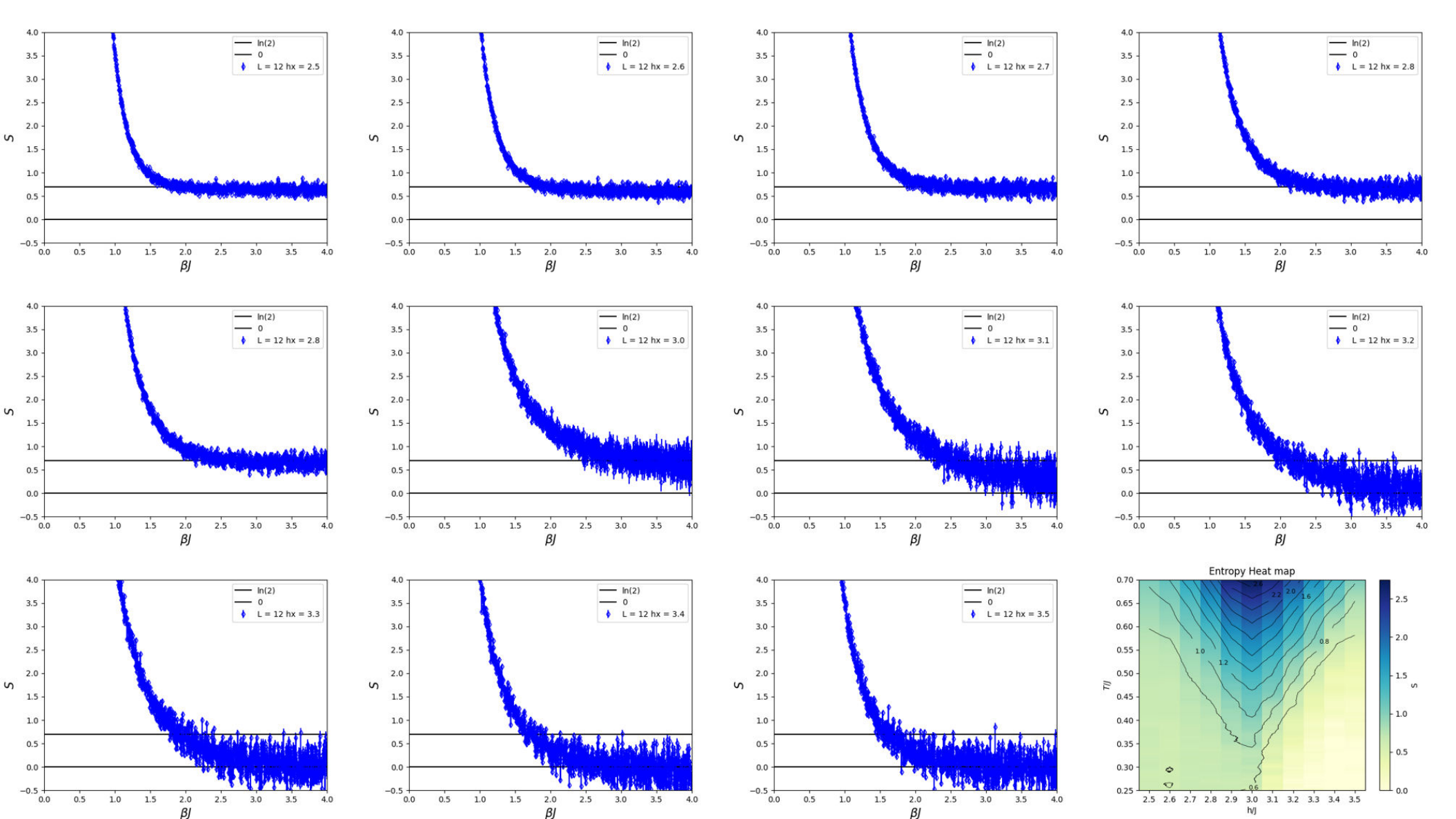}}\\
    \subfloat{\includegraphics[width=\textwidth]{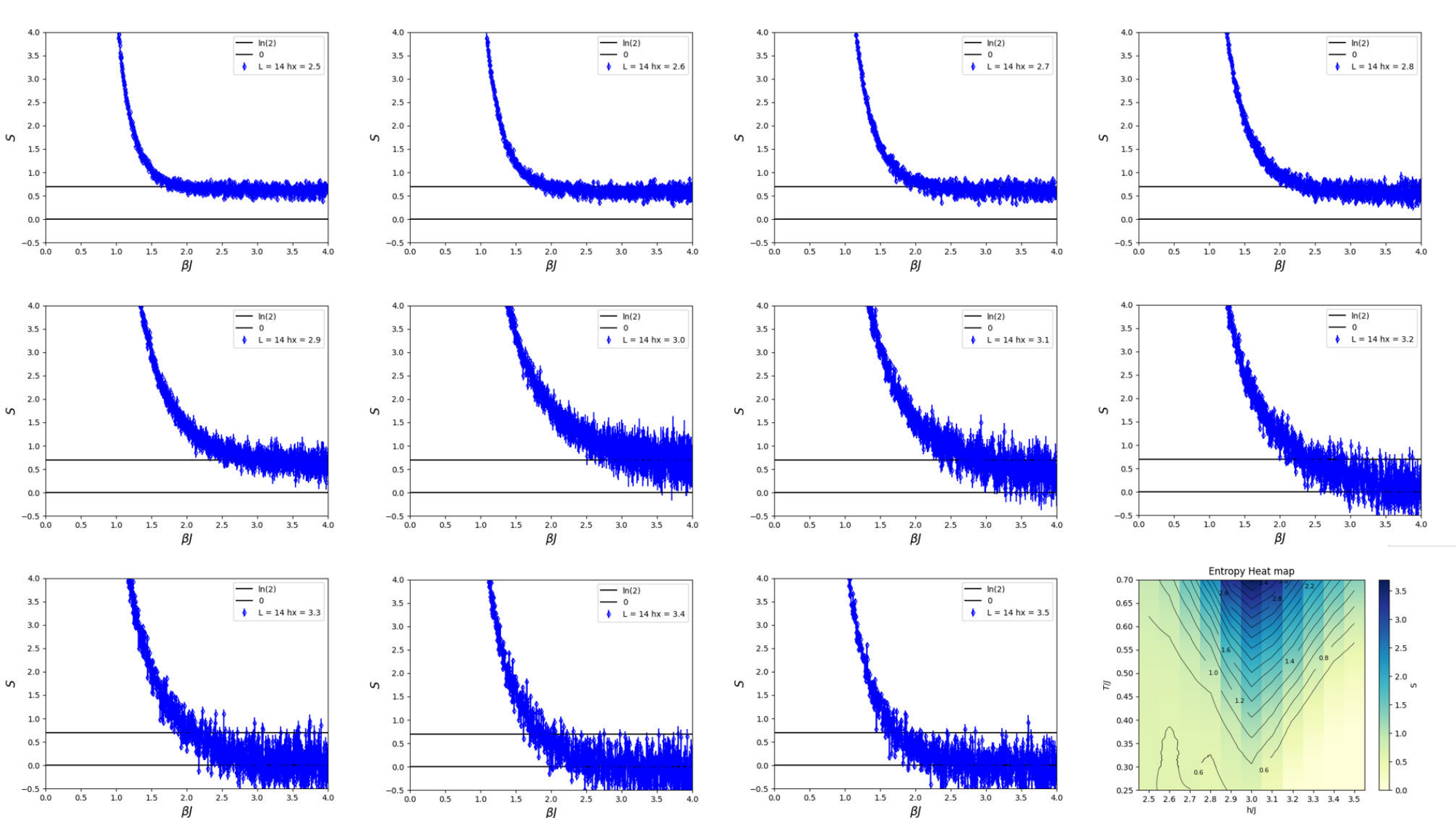}}
\end{figure} 

\clearpage

\begin{figure}[H]
  \centering
  \subfloat{\includegraphics[width=\textwidth]{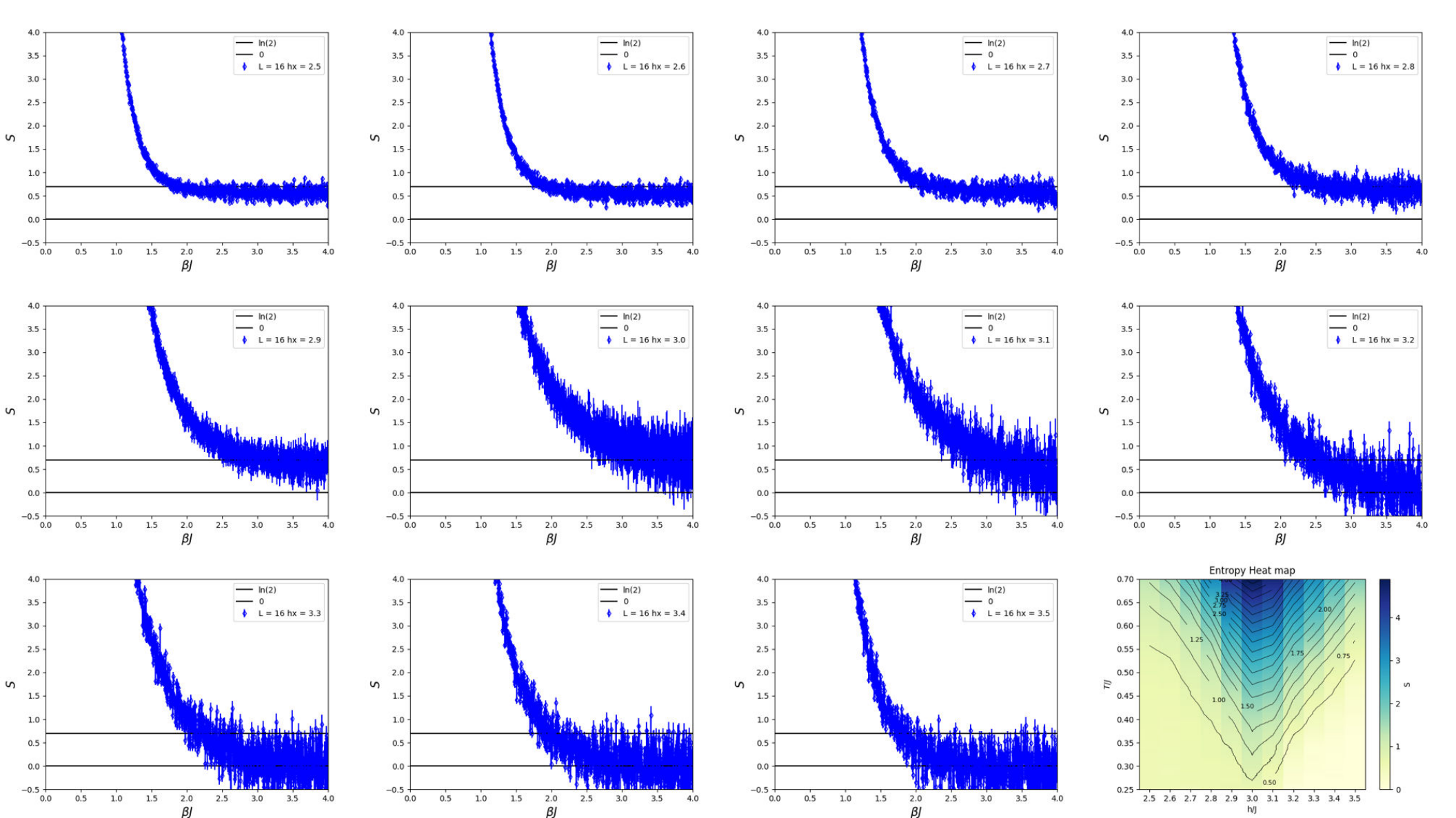}}
  \centering
  \caption{Entropy as function of inverse temperatures for $ L=12,14,16 $ square lattice quantum Ising model.}
  \label{fig:lg5}
\end{figure} 

\end{widetext}

\clearpage
As shown in \cref{fig:lg5}, we showed Monte Carlo entropy as a function of temperature for lattice sizes $ L = 12, 14,16 $. We take Trotter decomposition to be $\Delta_{\tau} = 0.01$. We take $ \beta  = 4 $ and cut into $ 1280 $ increments ($M=1280$). For each increment, we sample 1000 bins, for each bin, we perform $500$ combined updates as in \ref{sec:g1d}.

For the figure in the main text, we applied a smoothing procedure to the data by averaging the entropy values of neighboring points within an interval of $ \Delta \beta  = 0.05$.
\begin{widetext}

\begin{figure}[H]
  \centering
  \subfloat{\includegraphics[width=0.6\textwidth]{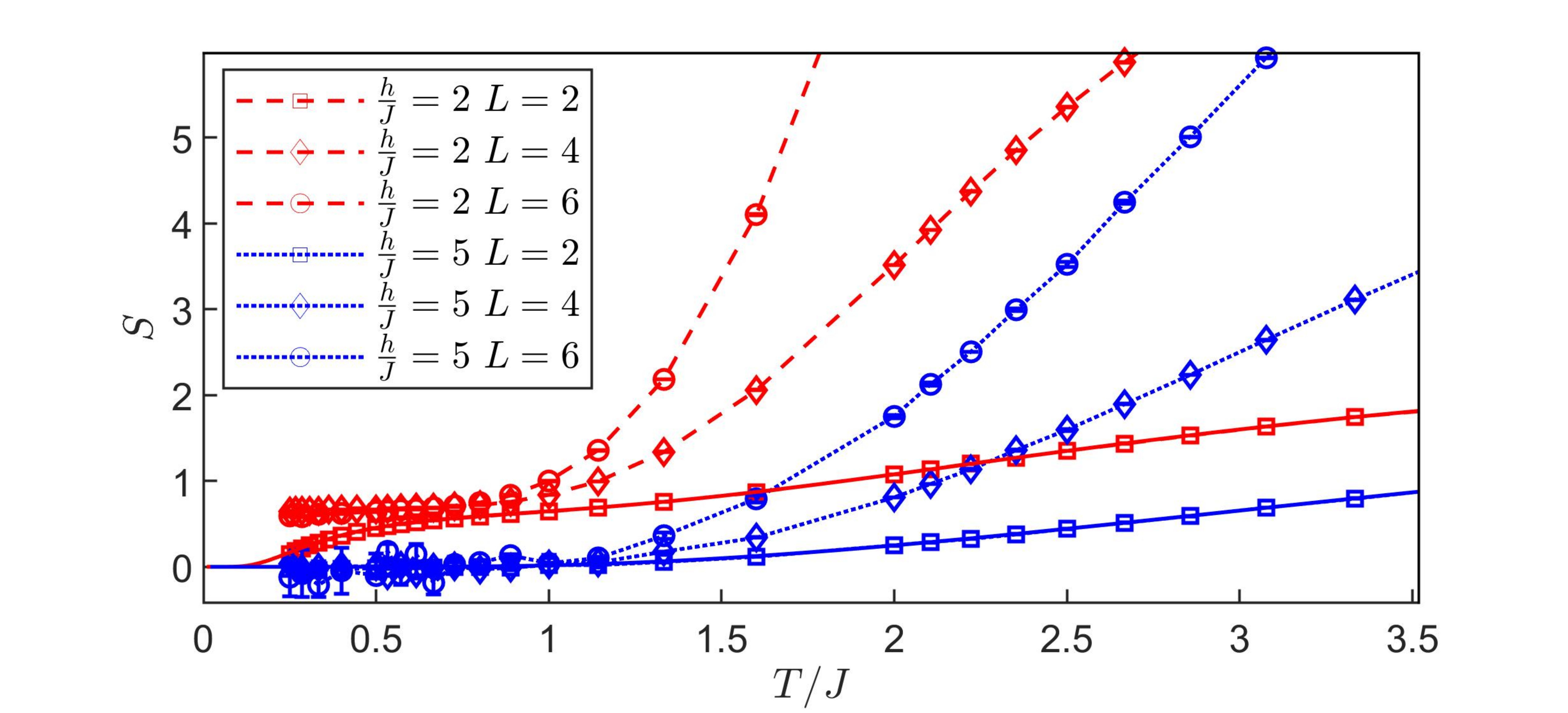}}
  \centering
  \caption{Entropy as function of temperature for $ L=2,4,6 $ square lattice quantum Ising model. Solid lines are results of exact diagonalization with $L=2$.}
  \label{fig:lgs}
\end{figure} 

\end{widetext}

We juxtaposed our TIMC outcomes with exact diagonalization results for a lattice size of $L = 2$, as illustrated in Fig.~\ref{fig:lgs}, which diminishes at much lower temperatures due to finite size effects—on a finite lattice, the model's ground state is invariably unique. As we reduce $h/J$ or increase the system size, the approximate $\ln 2$ plateau becomes more pronounced and extends closer to zero temperature, suggesting that it will persist in the thermodynamic limit.


\subsection{Triangular Lattice Quantum Ising model}

We also performed TIMC method on triangular lattice quantum lattice model with Hamiltonian given by 

\begin{equation}
\begin{aligned}
  H = -J \sum_{\langle i,j \rangle }\hat{Z}_{i} \hat{Z}_{j} - h \sum_{i}^{} \hat{X}_{i}
\end{aligned}
\label{}
\end{equation}
where we considered the antiferromagnetic case with $ J <0 $.

\begin{widetext}

\begin{figure}[H]
  \centering
  \subfloat{\includegraphics[width=\textwidth]{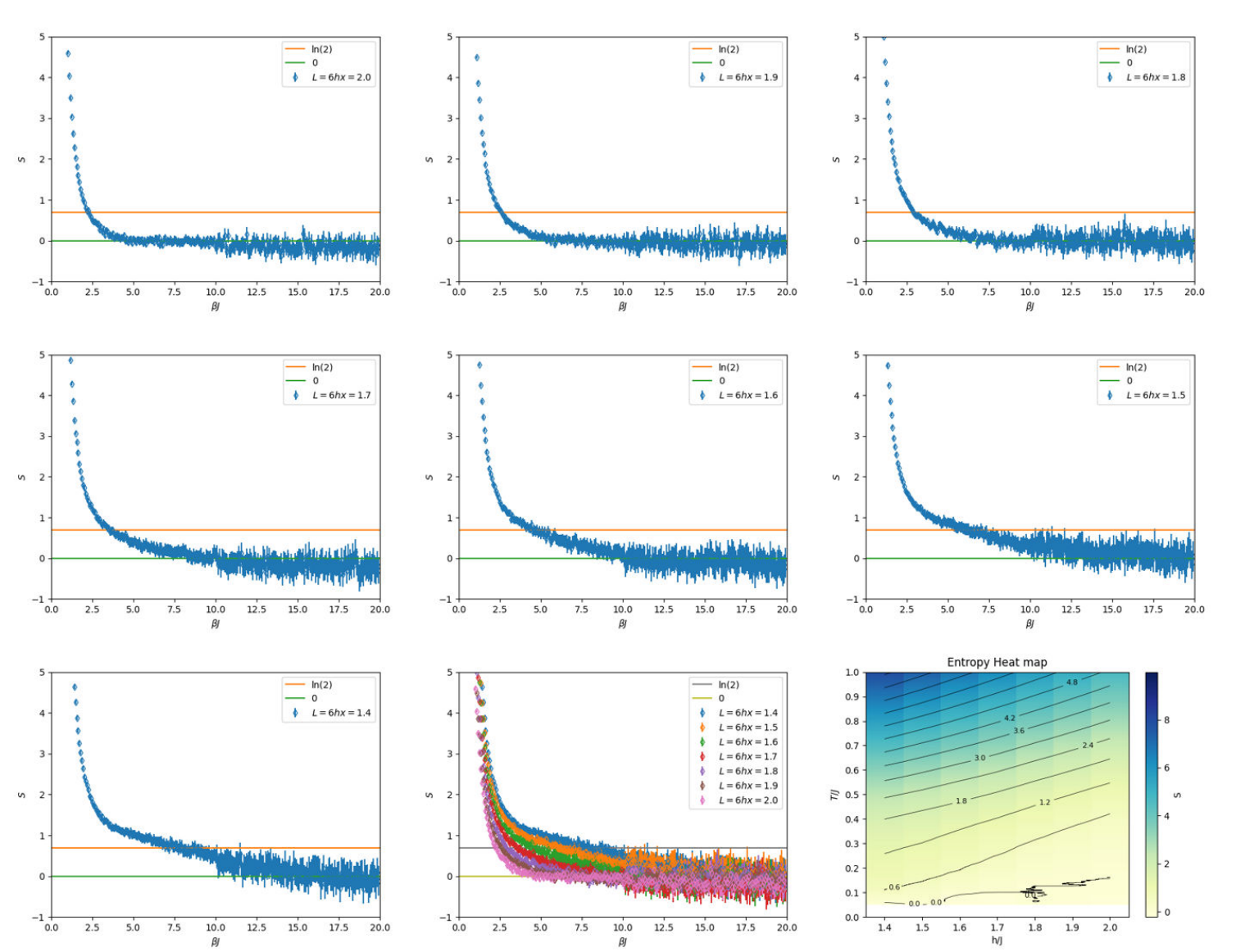}}\\
  \centering
  \caption{Entropy as function of inverse temperature for $ L=6 $ triangular lattice quantum Ising model.}
  \label{lgtri6}
\end{figure} 

\begin{figure}[H]
  \centering
  \subfloat{\includegraphics[width=\textwidth]{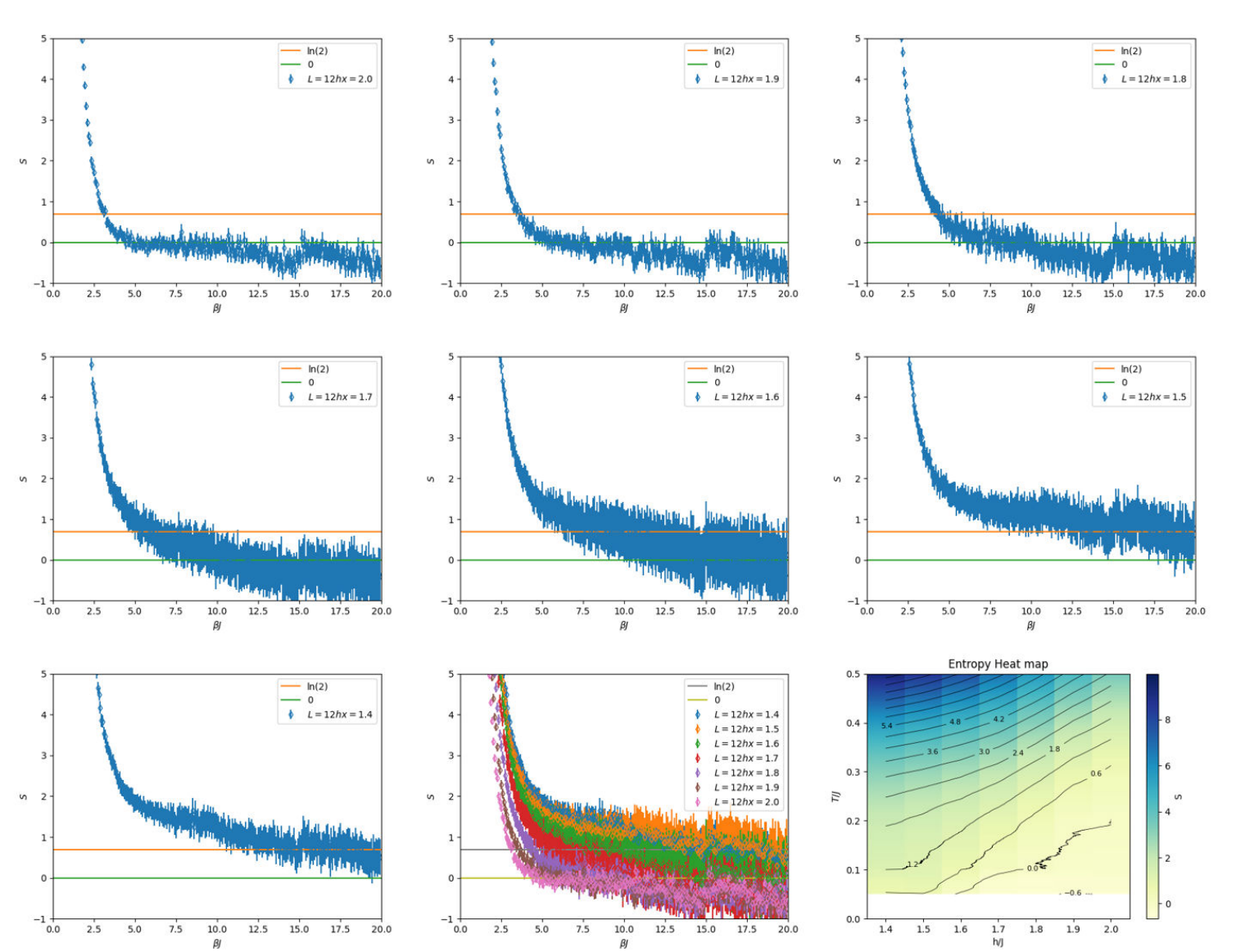}}\\
  \centering
  \caption{Entropy as function of inverse temperature for $ L=12 $ triangular lattice quantum Ising model.}
  \label{lgtri12s}
\end{figure} 
\end{widetext}

As shown in \cref{lgtri12s}, we showed Monte Carlo entropy as a function of inverse temperature for lattice scale $ L = 6,12$. We take Trotter decomposition to be $\Delta_{\tau} = 0.01$. In the region $ \beta  \in [0,10] $ we cut it into $ 1280 $ increments. In the region $ \beta \in [10,20] $ we cut it into $ 128 $ increments since the entropy is approaching a plateau. For each increment, we sample 1000 bins, for each bin, we perform $500$ combined updates as in the square lattice quantum Ising model.

\bibliographystyle{apalike}
\bibliography{Myref}

\end{document}